\documentclass[usenatbib,usegraphicx,useAMS]{mn2e}

\def\mgi{Mg\,{\sc i}}
\def\fei{Fe\,{\sc i}}
\def\sii{Si\,{\sc i}}
\def\tii{Ti\,{\sc i}}

\def\Teff{$T_{\rm eff}$}
\def\logg{$\log g$}
\def\logz{$\log Z$}
\def\um{$\mu$m}
\def\chisq{$\chi^{2}$}
\def\ga{\mathrel{\hbox{\rlap{\hbox{\lower4pt\hbox{$\sim$}}}\hbox{$>$}}}}
\def\la{\mathrel{\hbox{\rlap{\hbox{\lower4pt\hbox{$\sim$}}}\hbox{$<$}}}}

\def\msun{$M$\mbox{$_{\normalsize\odot}$}}

\def\minit{$M_{\rm init}$}
\def\lsun{$L$\mbox{$_{\normalsize\odot}$}}
\def\kms{\,km~s$^{-1}$}

\newcommand{\fig}[1]{Fig.\ \ref{#1}}

\title[RSGs as extra-galactic abundance probes]{The potential of Red
  Supergiants as extra-galactic abundance probes at low spectral
  resolution} \author[Davies, Kudritzki \& Figer]{Ben Davies$^{1,2}$,
  Rolf-Peter Kudritzki$^{3}$ and Donald F. Figer$^{1}$
  \\ $^{1}$Rochester
  Institute of Technology, 54 Lomb Memorial Drive, Rochester, NY
  14623, USA.  \\ $^{2}$School of Physics \& Astronomy, University of
  Leeds, Woodhouse Lane, Leeds LS2 9JT, UK \\ $^{3}$Institute for
  Astronomy, University of Hawaii, 2680 Woodlawn Drive, Honolulu, HI,
  96822, USA}
\begin{document}

\date{Accepted ... Received ...}

\pagerange{\pageref{firstpage}--\pageref{lastpage}} \pubyear{2009}

\maketitle

\label{firstpage}

\begin{abstract}
Red Supergiants (RSGs) are among the brightest stars in the local
universe, making them ideal candidates with which to probe the
properties of their host galaxies. However, current quantitative
spectroscopic techniques require spectral resolutions of R$\ga$17,000,
making observations of RSGs at distances greater than 1Mpc unfeasible.
Here we explore the potential of quantitative spectroscopic techniques
at much lower resolutions, $R \approx$2-3000. We take archival
$J$-band spectra of a sample of RSGs in the Solar neighbourhood. In
this spectral region the metallic lines of \fei, \mgi, \sii\ and
\tii\ are prominent, while the molecular absorption features of OH,
H$_2$O, CN and CO are weak. We compare these data with synthetic
spectra produced from the existing grid of model atmospheres from the
MARCS project, with the aim of deriving chemical abundances. We find
that all stars studied can be unambiguously fit by the models, and
model parameters of $\log g$, effective temperatures \Teff,
microturbulence and global metal content may be derived. We find that
the abundances derived for the stars are all very close to Solar and
have low dispersion, with an average of \logz=0.13$\pm$0.14. The
values of \Teff\ fit by the models are $\sim$150K cooler than the
stars' literature values for earlier spectral types when using the
Levesque et al. temperature scale, though we find that this
discrepancy may be reduced at spectral resolutions of $R=3000$ or
higher. In any case, the temperature discrepancy has very little
systematic effect on the derived abundances as the equivalent widths
(EWs) of the metallic lines are roughly constant across the full
temperature range of RSGs. Instead, elemental abundances are the
dominating factor in the EWs of the diagnostic lines. Our results
suggest that chemical abundance measurements of RSGs \textit{are}
possible at low- to medium-resolution, meaning that this technique is
a viable infrared-based alternative to measuring abundance trends in
external galaxies.

\end{abstract}

\begin{keywords}
stars: abundances, stars: late-type, stars: massive, supergiants,
galaxies: abundances, galaxies: stellar content
\end{keywords}

\begin{figure*}
  \centering
  \includegraphics[width=13.5cm,bb=0 0 550 850]{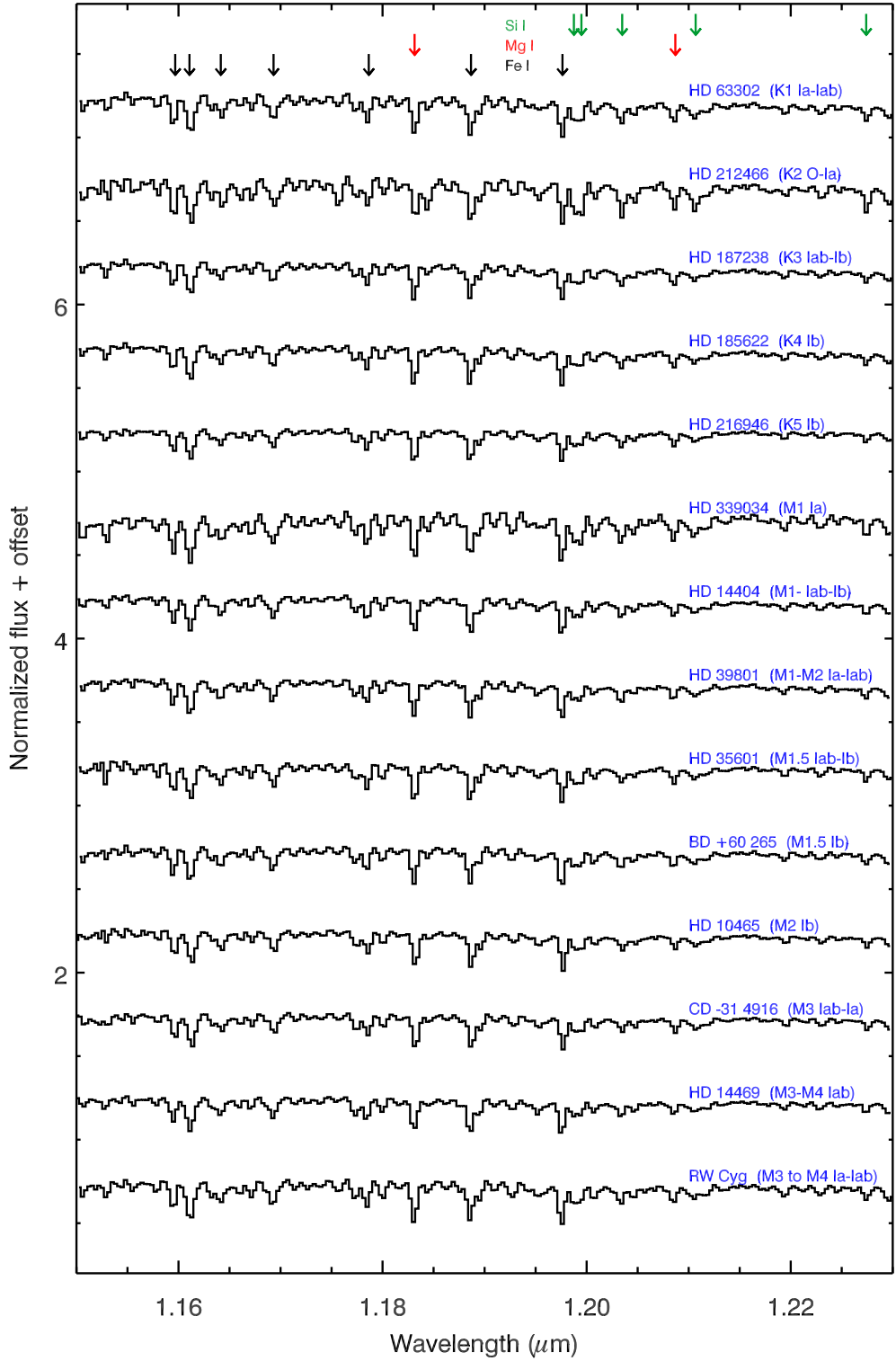}
  \caption{Spectra of RSGs in the $J$-band, at resolution of
    2,000. Positions and identifications of the dominant spectral
    features have been indicated. }
  \label{fig:specatlas}
\end{figure*}

\section{Introduction}
Red Supergiants (RSGs) are an evolved state of stars with initial
masses 8\msun$\la$\minit$\la$35\msun\ \citep[e.g.][]{Mey-Mae00}. They have
high bolometric luminosities ($L \ga 10^{4.5}$\lsun) which peak at
around 1\um, meaning that they are tremendously bright at optical
and near-IR wavelengths, with $M_{J}\sim -11$, rivalling the
integrated light of globular clusters and dwarf galaxies.

As the lifetime of massive stars is short, RSGs are necessarily young
objects, with ages of $\sim$50Myr at most. As they cannot have
travelled very far from their place of birth, the chemical composition
of RSGs should closely match that of the gas phase material in their
environs, aside from a certain amount of CNO processing. Therefore,
through quantititative spectroscopy of RSGs it should be possible to
use them as probes of their local chemical abundances.

The evolutionary predecessors of RSGs, blue supergiants of spectral
type A and B, have been very successfully used as tracers of chemical
abundances out to 7 Mpc distance by means of low resolution
quantitative spectral analyis at visual wavelengths
\citep{Kudritzki08,Kudritzki10}. Using individual stars in galaxies as
the source of information about chemical composition rather than
HII-regions and their emission lines has clear advantages, because the
latter are subject to substantial systematic uncertainties
\citep{K-E08,Bresolin09,Kudritzki10}. It is therefore important to
find as many reliable stellar abundances as possible useful for the
investigation of galaxies at larger distances. The goal of this paper
is to demonstrate that RSGs have an enormous potential in this regard.

A difficulty in extracting chemical abundances from RSG spectra is
that their cool atmospheres contain molecular material, and therefore
their spectra are filled with literally thousands of strong molecular
spectral lines. These lines overlap and blend with the metallic
lines. To extract abundance information from the spectra it is
necessary to observe at sufficiently high resolution so as to separate
the molcular lines and isolate the metallic features. This typically
means spectral resolutions in excess of $R \ga$ 17,000. Obtaining high
signal-to-noise (SNR) data at this resolution out to Mpc distances is
extremely challenging, even for the largest telescopes.

Abundance studies of RSGs to date have typically concentrated on
high-resolution observations in the $H$-band, where there is a rich
concentration of both molecular and metallic lines
\citep[see][]{Cunha07,RSGGC,RSGCabund}. However, if one shifts focus
to the $J$-band, one sees that the molecular transitions are much
reduced in both number and in strength, with the dominant spectral
features being the metallic lines of \fei\ and the $\alpha$-elements
\mgi, \sii\ and \tii. Hence, it should be possible to observe this
spectral region at much lower spectral resolution while still being
able to extract abundance information. The only drawback is that one
loses information about stellar nucleosynthesis within the RSGs
themselves, which is redundant information if one is only concerned
with the abundances of the local interstellar medium. This method of
deriving chemical abundances from $J$-band spectra is largely untested
and unexplored, the only such study to date being that of
\citet{Origlia04}.

In this paper we will explore the potential of quantitative
spectroscopy of RSGs in the $J$-band, using archival data and current
grids of model atmospheres. In Sect.\ \ref{sec:danda} we will describe
the $J$-band observations of RSGs, the model atmospheres, and the
fitting techniques we employ. The results of our analysis are
presented and discussed in Sect.\ \ref{sec:res}. We conclude in
Sect.\ \ref{sec:conc}.


\section{Data and analysis} \label{sec:danda}

\subsection{$J$-band observations of RSGs}
For the majority of our analysis we use data from the IRTF spectral
library. The spectra cover the wavelength range 0.8-4\um, at spectral
resolution of $R=2,000$, and at SNR$\ga$100. The spectra are all of
bright stars in the Solar neighbourhood. For full details of the
observations, see \citet{Rayner09}.

In \fig{fig:specatlas} we plot these spectra, in order of increasing
spectral type, in the wavelength range of 1.15-1.23\um. The spectra
have been normalized by the continuum flux. This continuum level is
not trivial to measure in spectra with many features -- we measured it
by ranking the channel values in the plotted range in order of
decreasing flux level, and then taking the median of the ten highest
values. The prominant spectral features have been identified,
following \citet{Wallace00}.

The first thing to note is that the dominant spectral features at this
resolution are due to metallic lines of \fei, \mgi, \sii\ and \tii. The
fluctuations in continuum level are not due to noise, these are blends
of molecular lines. However, in this wavelength range and at this
resolution, the lines are very weak and behave as a
`pseudo'-continuum. The second thing to note is that there is very
little variation between spectral types; from K0 to M4 the lines
appear to have very similar strengths. This may be due in part to the
narrow temperature range spanned by RSGs ($\sim$3500-4100K), and the
fact that all metal atoms are in the lowest ionization state at these
temperatures.

\subsection{Model atmospheres}
For this study we use the existing model grids of the MARCS project
\citep{Gustafsson08}, which are computed in LTE with spherical
symmetry for a range of stellar parameters. The mass of the star is
kept as a free parameter in computing the hydrostatic structure of the
spherically-symmetric atmosphere, and in this grid only masses up to
5\msun\ are computed. As RSGs are typically thought to have masses of
15-20\msun, our use of the 5\msun\ masses must be justified. The
effect of the stellar mass in the MARCS model atmospheres is to
control the so-called `extension' (or geometrical thickness) of the
atmosphere, $z$ \citep[for an explanation of this quantity
  see][]{Gustafsson08}. Since $z \propto L^{0.5}/M$, and $z \sim 20$
in Solar units for both Red Giants and RSGs, one does not expect there
to be much difference between the model spectra of stars with masses
2-20\msun\ if all other parameters are the same. Indeed, we find very
little difference between the spectra of models with 2\msun\ and
5\msun.


One caveat with the use of MARCS models for RSGs is that the
atmospheres of RSGs are known to be inhomogeneous and out of
hydrostatic equilibrium. Convective cells with sizes comparable to the
stellar radii and with radial velocities greater than the local sound
speed are thought to permeate the star. This can lead to a patchy
temperature structure at the surface and atmospheres that are extended
(i.e. greater pressure scale heights) compared to that predicted by
hydrostatic equilibrium \citep[see
  e.g.][]{Schwarzschild75,J-P07,Chiavassa09}. However, as the
scale-height is inversly proportional to gravity, we may expect a
model with increased scale-height to be the equivalent of a model with
lower gravity. Also, as we will show later the metallic lines in the
spectra of RSGs are very insensitive to temperature. Therefore, we
expect that the effects of patchy surface temperature and increased
scale-heights will have negligible impact on the derived abundance
levels.


We also note that the spectra in the MARCS model grid are not strictly
spectra, they are samplings of the SED flux at a grid of
wavelengths. As the flux between the sampling points is not known, if
any features are `missed' this can introduce artifacts into the
spectrum when degraded to lower resolution \citep[for a discussion of
  this effect see][]{Plez08}. We have run detailed comparisons between
the MARCS flux-sampled `spectra' and the fully-computed
high-resolution synthetic spectrum from MARCS model atmospheres
obtained from the {\sc pollux} database\footnote{\tt
  http://pollux.graal.univ-montp2.fr}, and we find that differences
are minimal when the spectra are degraded to lower resolution; the
only noticeable systematic difference is at the blend of \fei\ and
\sii\ at 1.198\um. In future studies we will use fully computed
spectra as they become available, but for now we proceed with the
existing grid.

In the MARCS grid effective temperatures are sampled at \Teff/K =
[3200,3400,3600,3800,4000,5000], which covers the entire range
spanned by RSGs. Metallicities are sampled at \logz=-1.5, and between
-1.0 and +1.0 in steps of 0.25dex (normalized to Solar
metallicity). This is easily sufficient to cover all metallicities
expected in the Galaxy. The sampling in microturbulence is poorer,
only values of $\xi$/\kms=[2,5] are computed. As a crude way of
improving this sampling, we have linearly interpolated the spectra
between these models to estimate the spectra of models with
$\xi$=3\kms\ and 4\kms. The sampling in \logg-space is also poor for
these models. For now we use only the models with \logg=(0.0, +1.0)
(cgs units), which have been shown to be typical values for RSGs
\citep[see e.g.][]{RSGCabund}.

The MARCS models are also available with different CNO abundance
mixtures. The grid sampling is complete for the unaltered case
(i.e. relative abundances of CNO set to Solar mixture). However, RSGs
are known to display the results of CNO processing at their surfaces
\citep[i.e. enhanced N, depleted C, see][]{RSGCabund}. For this reason
we also examine the MARCS models with CN processing, though the grid
sampling is poorer for these models.

\begin{figure}
  \centering
  \includegraphics[width=9cm]{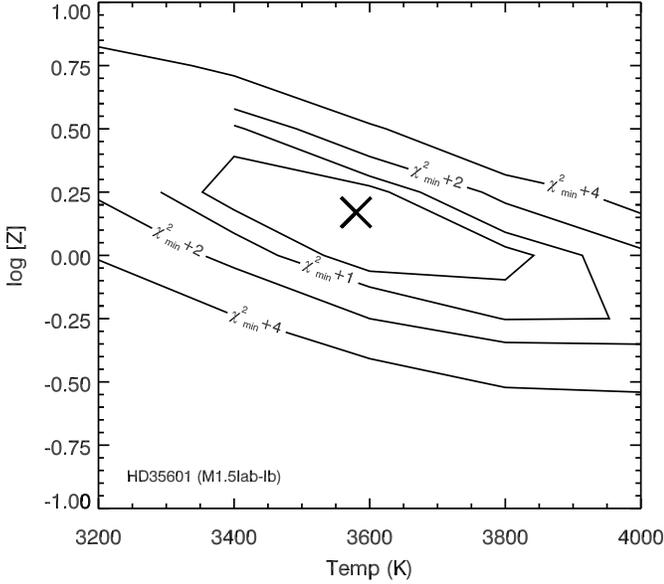}
  \caption{An example of the fitting process for CD-31 49. The figure
    shows contours of equal \chisq\ in the \logz-\Teff\ plane, at
    values of 0.5, 1.0, 2.0 and 4.0 above the minimum \chisq. The
    cross indicates the location of the \chisq\ minimum. The gap in
    the contours is due to a missing model in the grid at \Teff=3200K,
    \logz=+0.5.  }
  \label{fig:example_chisq}
\end{figure}

\begin{figure*}
  \centering
  \includegraphics[width=15cm]{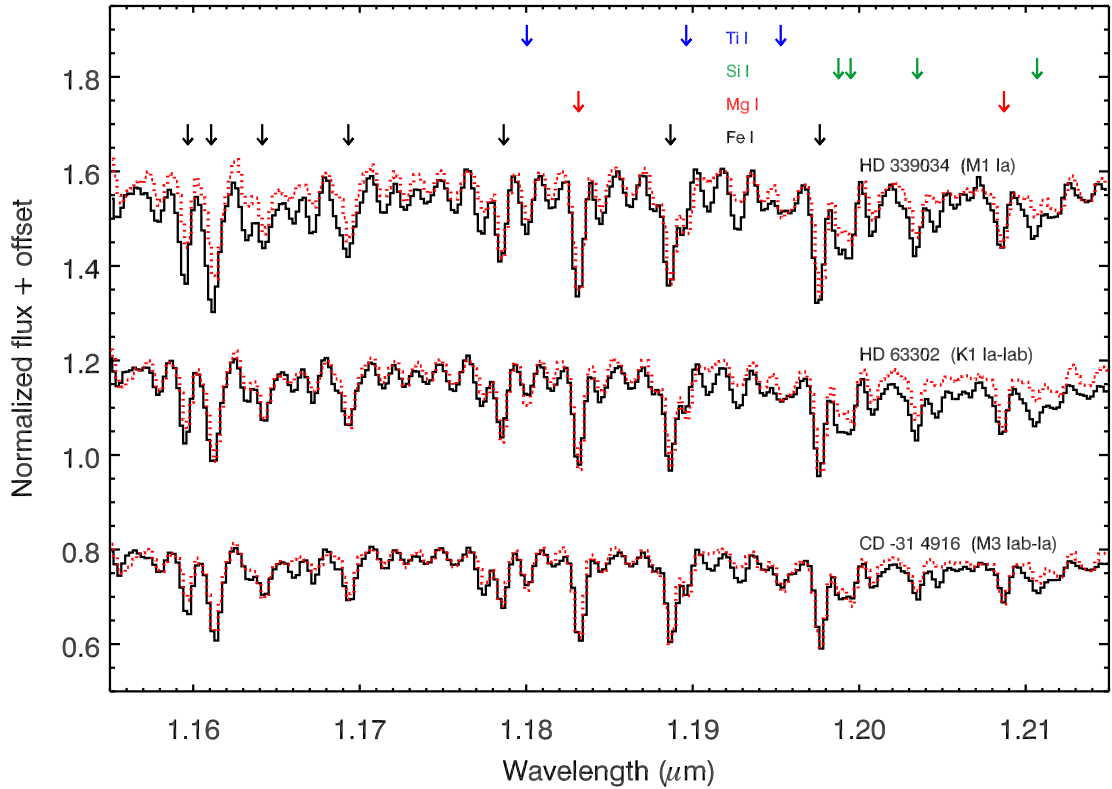}
  \caption{Examples of model fits (red dotted lines) to the data
    (black solid lines). Data have been resampled onto a 2$\times$
    finer grid to help aid the identification of discrete spectral
    features. The top spectrum is an example of a poor fit, with
    reduced \chisq=8.9. The middle spectrum is
    a ``fair'' fit, with \chisq=4.5. The bottom spectrum has a good
    fit, with \chisq=1.7. }
  \label{fig:example}
\end{figure*}

\subsection{Fitting procedure} \label{sec:fit}
Before comparing the observed and model spectra, we first degrade the
spectral resolution of the model data to that of the observations
using a gaussian convolution kernel. Then, in order to find the
best-fit spectrum, we have experimented with several
strategies. Firstly, we compare the normalized flux at each spectral
channel and compute the reduced \chisq\ for each point in the model
grid,

\begin{equation}
\chi^{2} = \frac{1}{\nu}  \sum \displaystyle\left( 
\frac{f_{i, \rm obs} - f_{i, \rm mod}}{\sigma_{i}}
\right)^{2}
\label{equ:chisq}
\end{equation}

\noindent where $f_{i, \rm obs}$ and $f_{i, \rm mod}$ are the observed
and model fluxes at each wavelength channel $i$; $\sigma_{i}$ is the
uncertainty in the flux, taken to be 0.01 at each channel
(i.e. SNR=100); and $\nu$ is the number of degrees of freedom,

\begin{equation}
\nu = N - n - 1
\label{equ:nu}
\end{equation}

\noindent where $N$ is the number of wavelength channels, and $n$ is
the number of model parameters being fitted. In the case of the MARCS
grid, $n=3$ (\logz, \Teff, and $\xi$). We then take the best fitting
model to be that point in the grid with the lowest value of \chisq.

A second fitting method applies a similar principle, but this time
instead of fitting the entire spectral region we fit only the
equivalent widths (EWs) of the strongest spectral lines: those of
\fei\ (1.16108\um, 1.16414\um, 1.18866\um, 1.19763\um) and
\mgi\ (1.18314\um, 1.20869\um). We again find the best fitting model
by determining the point in the model grid with the lowest \chisq. We
estimate the typical uncertainty in EW measurements ($\sigma_{i}$) to
be around 10\%, from repeating EW measurements of lines and slightly
varying the wavelength ranges.

The first `full-spectrum' method has the advantage that it fits the
whole spectrum, and so in principle can deal with blends of lines. The
second `strong-lines' method on the other hand takes into account only
the most prominent spectral features and so is not swayed by
unidentified blends for which the physical data may be poorly known. In
practice, we settled on a method which is a hybrid of these two -- we
look at the residuals between the model and observed spectrum, as in
the `full-spectrum' method, but only at select regions of the
spectrum. This allows us to place more emphasis on the strong lines,
but also deal with blended lines such as the \fei-\sii\ blend at
1.198\um. 

To determine the best-fit physical properties and their respective
uncertainties, we first find the point in the model grid with the
lowest \chisq, \chisq$_{\rm best}$. We then take into account all
points in the grid with \chisq-\chisq$_{\rm best} < \Delta$\chisq,
where $\Delta$\chisq\ is an arbitrary number. If there are fewer than
10 models within this range, we take the 10 models with the best
\chisq. We then compute the weighted mean of a physical property
$\bar{X}$ and its uncertainty $\sigma \bar{X}$ from the relations,

\begin{equation}
\bar{X} = \frac{\sum^{N}_{i} w_{i} x_{i}}
{\sum^{N}_{i} w_{i}} , ~~~
\sigma\bar{X} = \sqrt{ \frac{N'}{N'-1}
\frac{ \sum^{N}_{i} w_{i} (x_{i} - \bar{X})^2 } {\sum^{N}_{i} w_{i}} }
\label{equ:mean}
\end{equation}

\noindent where $x_i$ is that physical property (e.g. \Teff, $\xi$,
\logz) in model $i$, and the weights $w_{i}$ are calculated from the
\chisq\ of each model in the grid,

\begin{equation}
w_{i} = e^{-\chi^2/2}
\label{equ:weights}
\end{equation}

\noindent and $N'$ is the number of non-zero weights. The parameter
which governs how many models are used in the computation of the
physical parameters, $\Delta$\chisq, is initially set to 2 (see also
Mottram et al., submitted, for similar use of this analysis
technique). However, we found that this significantly overestimated
the errors, particularly in \logz\ (as will be discussed in
Sect.\ \ref{sec:res}), and we concluded that $\Delta$\chisq=1 was more
appropriate.

An example of this process is shown in \fig{fig:example_chisq}, where
we show \chisq\ contours in the \Teff-\logz\ plane. The cross
indicates the derived temperature and metallicity for one star in the
sample. It can be seen from the figure that the inferred \logz\ has
only a weak sensitivity to the fitted \Teff.

\begin{table*}
  \centering
  \caption{Derived parameters for the programme stars. Temperatures
    are those appropriate for their spectral types from the
    calibration of \citet{Levesque05}; where a star is between
    spectral types we linearly interpolate between the two. Abundance
    types are either `st' for standard, or `CN' for CN processed.  }
  \begin{tabular}{llcccccc}
\hline
\hline
      Name  &  Spec Type  &  T(K) &  Tfit(K) & log[Z] & $\xi$/\kms & AbT &   \chisq \\
\hline
   HD63302 &   K1Ia-Iab & 4100 & 3660 $\pm$ 180 &  0.02 $\pm$ 0.20 & 3.4 $\pm$ 0.5 & st &   4.47 \\
  HD212466 &     K2O-Ia & 4015 & 3770 $\pm$ 170 &  0.17 $\pm$ 0.20 & 3.8 $\pm$ 0.5 & st &   5.57 \\
  HD187238 &   K3Iab-Ib & 4015 & 3630 $\pm$ 180 &  0.04 $\pm$ 0.21 & 2.3 $\pm$ 0.5 & st &   4.91 \\
 HD185622A &       K4Ib & 3928 & 3590 $\pm$ 170 &  0.12 $\pm$ 0.20 & 2.4 $\pm$ 0.5 & st &   5.95 \\
  HD216946 &       K5Ib & 3840 & 3580 $\pm$ 230 & -0.15 $\pm$ 0.27 & 2.5 $\pm$ 0.5 & CN &   3.29 \\
   HD14404 &  M1-Iab-Ib & 3745 & 3580 $\pm$ 170 &  0.10 $\pm$ 0.19 & 2.5 $\pm$ 0.5 & CN &   4.45 \\
   HD39801 & M1-M2Ia-Ia & 3710 & 3520 $\pm$ 160 &  0.19 $\pm$ 0.21 & 2.3 $\pm$ 0.5 & st &   6.59 \\
   HD35601 & M1.5Iab-Ib & 3710 & 3630 $\pm$ 160 &  0.11 $\pm$ 0.19 & 2.8 $\pm$ 0.5 & st &   3.87 \\
 BD+60 265 &     M1.5Ib & 3710 & 3570 $\pm$ 140 &  0.16 $\pm$ 0.17 & 2.3 $\pm$ 0.5 & st &   6.90 \\
  HD339034 &       M1Ia & 3745 & 3820 $\pm$ 160 &  0.29 $\pm$ 0.15 & 3.5 $\pm$ 0.5 & st &   8.86 \\
   HD10465 &       M2Ib & 3660 & 3520 $\pm$ 220 & -0.08 $\pm$ 0.27 & 3.4 $\pm$ 0.6 & st &   1.78 \\
   HD14469 &   M3-M4Iab & 3550 & 3530 $\pm$ 160 &  0.15 $\pm$ 0.22 & 2.4 $\pm$ 0.5 & CN &   5.76 \\
  CD-31 49 &   M3Iab-Ia & 3605 & 3580 $\pm$ 220 & -0.15 $\pm$ 0.29 & 3.1 $\pm$ 0.6 & st &   1.67 \\
    RW Cyg & M3toM4Ia-I & 3550 & 3680 $\pm$ 180 &  0.31 $\pm$ 0.21 & 2.6 $\pm$ 0.5 & st &   6.98 \\
\hline
    \label{tab:averesults}
  \end{tabular}
\end{table*}

As an additional check on the significance of each abundance
measurement, we look at the \logz\ of the models in the grid with
\Teff$\pm$200K and $\xi \pm$1\kms. In each extreme, we determine the
`best-fit' \logz\ from the model with the lowest \chisq. Therefore, we
derive a range of \logz\ values by varying \Teff\ and $\xi$. In the
vast majority of cases, we find that the range in \logz\ is less than
the sampling of the model grid (0.25dex), and in the worst case two
grid points ($\pm$0.25dex). This is smaller than the typical
uncertainties on each measurement (see next Section).


\section{Results} \label{sec:res}

Before moving on to the derived stellar parameters for the programme
stars, we first give examples of the quality of fits we achieve. In
\fig{fig:example} we show the range of fits obtained, from the worst
to the best. In the case of the worst fit (HD~339034, top spectrum in
\fig{fig:example}), it is clear that in the most part the strongest
lines, those of \fei\ and \mgi, are still matched very well, apart
from at the blue end of the displayed spectrum where there may be a
problem with the continuum normalization. The deficiencies come mainly
from unresolved and unidentified blends, such as that at 1.194\um, and
from the \sii\ lines. However, at higher spectral resolution the \sii\ lines
would be unblended and could be incorporated into the fitting
procedure, possibly obtaining much better fits (as will be shown in
Sect.\ \ref{sec:specres}). A thorough examination of this requires
higher resolution data and is left for a future study.

The other stars in \fig{fig:example} are fit very well by the
procedure. This quality of fit is more typical of the results, as can
be seen from the last column of Table \ref{tab:averesults}.

\begin{figure}
  \centering
  \includegraphics[width=9cm]{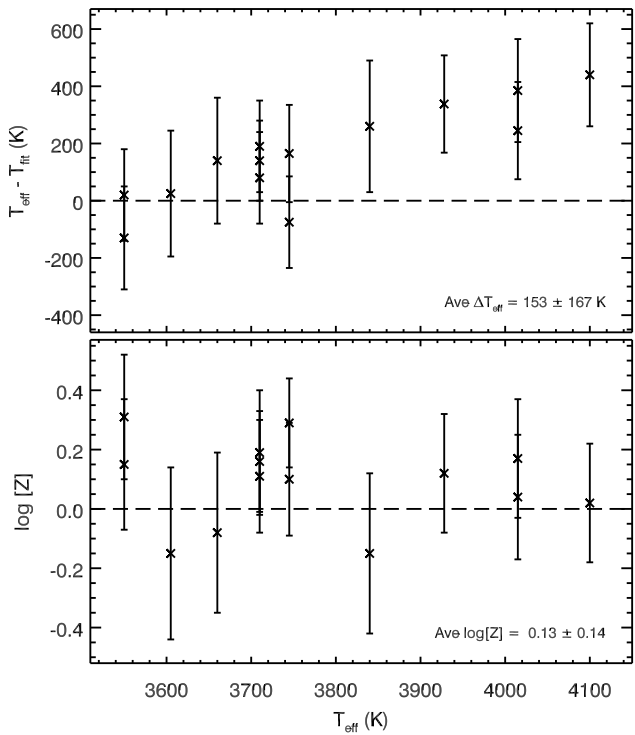}
  \caption{Results of the analysis process on the programme stars, as
    a function of their temperatures inferred from their literature
    spectral types and the temperature calibration scale of
    \citet{Levesque05}. The top panel shows the difference in
    \Teff\ between the literature values and that derived by the
    fitting procedure. The bottom panel shows the derived average
    metal content \logz. The weighted means of each parameter are
    shown in the panels. }
  \label{fig:averesults}
\end{figure}

\subsection{Derived stellar parameters}

We now address the results of the fitting procedure across all stars
studied. The temperatures, microturbulent velocities and average
abundances for each star are listed in Table \ref{tab:averesults}. In
\fig{fig:averesults} we show the values of \Teff\ and \logz\ derived
for each star, as well as the averages of each parameter, as a
function of the stars' literature temperatures, assuming the
temperature scale of \citet{Levesque05}. In terms of temperature, we
find generally acceptable agreement with the literature values. We do
find on average that the temperatures we derive are cooler by
$\approx$150K, with a 1$\sigma$ standard deviation of 170K on the
distribution. This effect is more pronounced at higher
temperatures. However, if one takes the old temperature scale from
\citet{H-McE84}, which are on average $\sim$200K cooler than those of
Levesque et al., the agreement is almost perfect. This is curious, as
the RSG temperature scale was rederived from the current generation of
MARCS model atmospheres, but concentrating mainly on the $R$ and $I$
spectral windows. We will discuss how the derived temperature may be
affected by line blending in Sect.\ \ref{sec:disc}, but we will also
show in Sect.\ \ref{sec:robust} that the fitted stellar temperature
has very little impact on the measured abundances, so for now we are
not concerned by this minor discrepancy.

The fitted values of \logz\ are very tightly distributed around zero,
with mean metallicity of \logz=$0.13 \pm 0.14$, where the uncertainty
again is the 1$\sigma$ standard deviation. These values are therefore
consistent with Solar metal abundances, which is a reassuring result
for a sample of objects in the Solar neighbourhood. If we assume for
the moment that all objects in the sample have \logz=0.0, then the
bottom panel of \fig{fig:averesults} serves to illustrate an {\it
  empirical} determination of the precision of our analysis method,
that is that we are deriving abundances accurate to
$\pm$0.14dex. However, there is likely some intrinsic spread in the
metallicities of the objects in our sample, so this value is an upper
limit to the experimental uncertainty. This compares to the average
errors we determined in the fitting process of $\sim$0.2dex (see
column 4 of Table \ref{tab:averesults}). So, it is likely that our
analysis process overestimates the experimental errors slightly. When
we used $\Delta$\chisq=2 in the fitting process (see previous Section)
these errors were higher still, at around $\pm$0.25-0.3dex, yet the
average \logz\ and its standard deviation were the same as with
$\Delta$\chisq=1. This is the reason we opted for $\Delta$\chisq=1
during the fitting, since the errors were more comparable to
`empirical' error of $\pm$0.14dex. This level of precision is as good
as any other abundance measurement technique.

For the microturbulent velocities $\xi$, all measured values are
between 2-4\kms, with an average of $2.8 \pm 0.5$\kms. This is a
typical value for RSGs \citep[see e.g.][]{RSGCabund,Cunha07}. We note
however that the $\xi$-sampling of the MARCS grid is very coarse, with
computed models at 2 and 5\kms\ only. Our grid points at 3 and
4\kms\ were simple linear interpolations in between the computed
models, and so if non-linear effects are present there will be a
systematic uncertainty in our derived values of $\xi$. However, we can
say that the current $\xi$-sampling of the model grid encompasses the
range of values expected from the RSGs in our sample. The impact of
the uncertainty in $\xi$ on abundance measurements will be explored in
the next Section.

We explored two sets of models with different CNO mixtures -- the
`standard' set, and the CNO-processed set with depleted C and enhanced
N. For the majority of the objects studied the standard CNO abundance
models provided the best fits. As with $\xi$, our sensitivity to this
free parameter will be explored next.

\begin{figure}
  \centering
  \includegraphics[width=9cm]{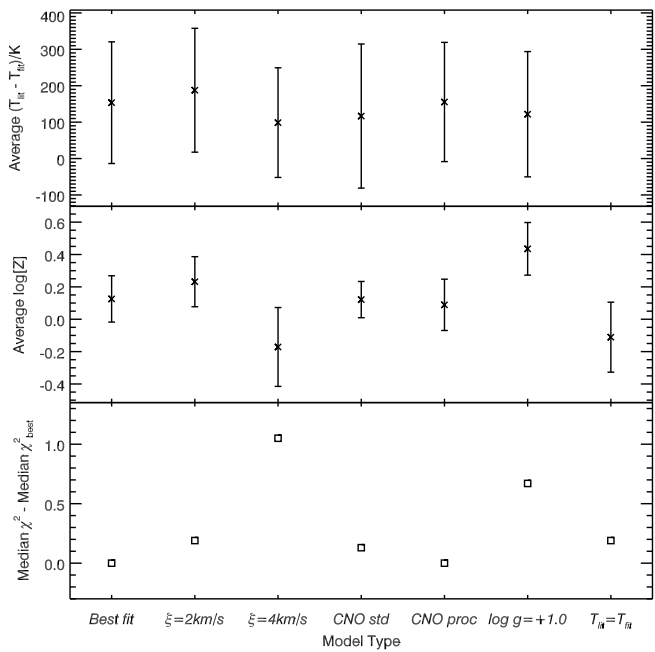}
  \caption{The impact of fixing various input parameters on the
    results of the fitting process. Each panel shows the results of
    the best fit (i.e. letting all parameters vary), fixing
    $\xi$=2\kms\ and $\xi$=4\kms, fixing the CNO elemental abundances
    to the `standard' and `processed' values, and fixing
    \logg=+1.0. The top panel shows the weighted-mean abundance level
    of all stars and its standard deviation. The bottom panel shows
    the median \chisq\ of all fits from each of the fitting methods. }
  \label{fig:robust}
\end{figure}

\subsection{Robustness of abundance measurements} \label{sec:robust}
In the previous Section we have shown that the abundances we measure in
this technique are extremely consistent with the expected values, with
a very high level of precision. To investigate the robustness of these
measurements, we now explore the effect of varying the free parameters
on the results. In \fig{fig:robust} we plot the weighted mean of all
stars' fitted abundance levels, the standard deviation on this value,
and the reduced \chisq\ values when forcing each of the model input
parameters to be fixed. 

\paragraph*{\bf Sensitivity to microturbulence}
This parameter affects the EWs of strong lines. If lines are
saturated, then by adding a degree of microturbulence the overall
capacity to absorb is increased, and stronger absorption lines are
observed. The second and third points on the $y$-axis in
\fig{fig:robust} show the impact of forcing the fitting routine to
find the best fit when the microturbulence is fixed at
$\xi$=2\kms\ and $\xi$=4\kms. A systematic effect on the derived
\logz\ can be seen, that is for lower $\xi$, higher \logz\ is
found. However, the difference is small ($\pm$0.2dex), while larger
dispersion in \logz\ is found, illustrated by the size of the error
bar. The larger dispersion indicates that, on average, poorer fits are
found, in accordance with the larger values of \chisq\ from the
fitting procedure (bottom panel of \fig{fig:robust}). The small change
in the average \logz\ suggests that the uncertainties in abundance
levels propagated through from uncertainties in $\xi$ are typically
less than the experimental error in \logz.

\paragraph*{\bf Sensitivity to CNO abundances}
We find very little sensitivity to this input parameter. Whether the
`standard' or `CNO processed' abundances are used, the derived
\logz\ is very close to the best-fit value, with similar error-bar
(see the fourth and fifth points on the $y$-axis in
\fig{fig:robust}). 

\paragraph*{\bf Sensitivity to gravity}
It is not possible as yet to make as thorough an investigation of the
effect of varying this parameter as for the others, as it is poorly
sampled in the MARCS grid for supergiants. However, models do exist
for \logg=+1.0 and 0.0. These values probably bracket the accepted
values for most RSGs, with \logg=0.0 being the most commonly used
number \citep{Lambert84,Carr00,Levesque05,Cunha07}. Though values of
\logg\ as low as -0.5 have been measured for some RSGs, the vast
majority of RSGs have gravities of \logg=0.0 and higher \citep[see
  Fig.\ 5 of][]{Cunha07}. We find that, on average the \logg=+1.0 fits
produce significantly larger \chisq, and a slightly larger spread on
the \logz\ distribution. The average \logz\ however remains consistent
with the best-fit solution to within the errors. We experimented with
using the \logg=+1.0 models in the formal fitting procedure, and
deriving best-fit values for \logg\ rather than assuming a blanket
value of \logg=0.0. In all cases we found that the derived values of
\logg\ were between 0.0 and 0.2, with negligible effect on the derived
\logz\ and \Teff.

\paragraph*{\bf Sensitivity to input \Teff}
The top panel of \fig{fig:robust} illustrates that, for any set of
model parameters used, we consistently derive temperatures that are
systematically lower than the literature values, especially for
earlier spectral types. As commented upon earlier, this may be due in
part to the recently-rederived temperature scale for
RSGs. Nevertheless, to investigate the significance of this effect we
re-ran the fitting procedures but this time forced the fitted
temperature to be the same as the literature temperatures, allowing
the rest of the parameters to vary. The results are shown in the
right-most points of \fig{fig:robust}. Though the reduced
\chisq\ values are slightly worse than for any of the other model
runs, the average abundances are again extremely consistent with the
best-fit model, only with a slightly larger dispersion.

\begin{figure}
  \centering
  \includegraphics[width=9cm]{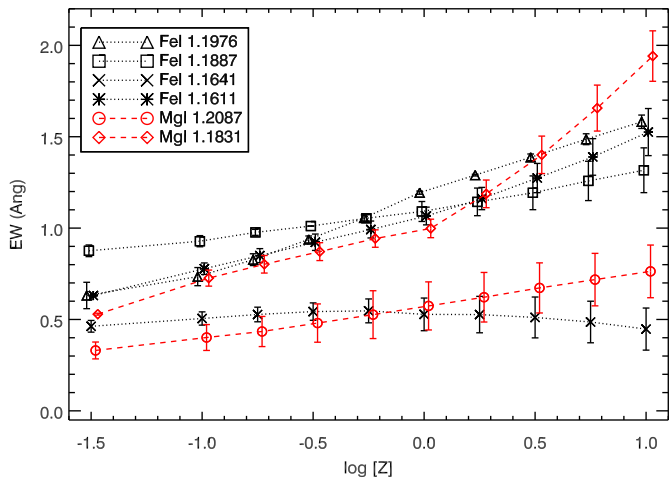}
  \caption{The effect of average metal content \logz\ on the EWs of
    the diagnostic lines used in this study, as measured from the
    synthetic spectra in the MARCS model grid. Points show the average
    EWs, while error bars show the range of EWs measured across the
    effective temperature range of RSGs (3400-4000K). All models use
    \logg=0.0, $\xi$=3\kms, and standard CNO abundances. The
    $x$-values of each of the curves have been slightly offset from
    one another for clarity. }
  \label{fig:zeffect}
\end{figure}

\smallskip

From this analysis, it appears that the free parameters the
fitting process, namely \Teff, $\xi$, \logg, and $X(\rm CNO)$, have
very little effect on the derived abundances. The standard deviation
on the results of each model run in \fig{fig:robust} is only 0.2dex,
with a mean of +0.0dex. These numbers are comparable to those derived
for the `best-fit' results. Thus, it seems that any systematic
uncertainty in the fitting process is only comparable to the random
error, and the derived abundance level is robust to fluctuations in
the model's free parameters.


\section{Discussion} \label{sec:disc}

\subsection{RSGs as abundance probes}

The results of this study suggest that the MARCS model atmospheres
provide excellent fits to the $J$-band spectra of Solar-neighbourhood
RSGs. Moreover, the derived abundances seem to be relatively
insensitive to the model input parameters \Teff, \logg, $\xi$ and the
mix of CNO abundances. Therefore, by constraining these input
parameters to the ranges that are known to be valid for RSGs, it
should be possible to derive metal abundances for RSGs to a precision
of 0.1dex using medium resolution $J$-band spectroscopy.

To illustrate the sensitivity of the $J$-band diagnostic lines to
average metal content, in \fig{fig:zeffect} we plot the strengths of
the strong lines identified in Sect.\ \ref{sec:fit} as a function of
\logz\ for the model spectra in the MARCS grid. The error-bars
indicate the minimum and maximum EWs observed across the range of RSG
temperatures, from 3400 to 4000K. All measurements were computed from
the \logg=0.0, $\xi$=3\kms\ models with standard CNO abundances. The
plot shows that, for these lines, there is very little variation with
RSG spectral type, presumably due to the narrow range of temperatures
that RSGs span. By contrast, the effect of changing metal content is
much more pronounced. For the typical abundance gradient in a spiral
galaxy of 0.05-0.2dex/kpc
\citep[see][]{Garnett97,Kudritzki08,Bresolin09,U09}, the strengths of
these lines could increase by $\sim$50\% between the innermost and
outermost regions. As such variations in line strength would be easily
detectable, RSGs could be powerful tools with which to study abundance
trends in their host galaxies.

\begin{table*}
  \centering
  \caption{Derived parameters for the two stars for which higher
    spectral resolution was available.}
  \begin{tabular}{llcccc}
    \hline
    \hline
    Name & Spec Type & $T$(K) & $T_{\rm fit}$(K) & $\log Z$ & $\xi$(\kms)\\
    \hline
$\alpha$ Her & M5~Ib    & 3450 & 3460$\pm$220 & -0.26$\pm$0.22 & 3.9$\pm$0.3 \\ 
$\alpha$ Ori & M1.5~Iab & 3710 & 3660$\pm$170 &  0.24$\pm$0.22 & 4.4$\pm$0.5 \\ 
    \hline
    \label{tab:r3000}
  \end{tabular}
\end{table*}

\subsection{Benefits of increased spectral resolution} \label{sec:specres}
The spectra studied here were taken from the IRTF spectral library,
the spectral resolution of which is $R=2000$. At this resolution we
have identified six spectral lines which have sufficient strength to
be largely unaffected by blending. Nevertheless, there is clearly some
blending present between the metallic lines, e.g. those of \fei\ and
\tii. At lower metallicities, where the metallic lines will tend to
have lower EWs, blending may begin to introduce problems. 

\begin{figure*}[t]
  \centering
  \includegraphics[width=17cm]{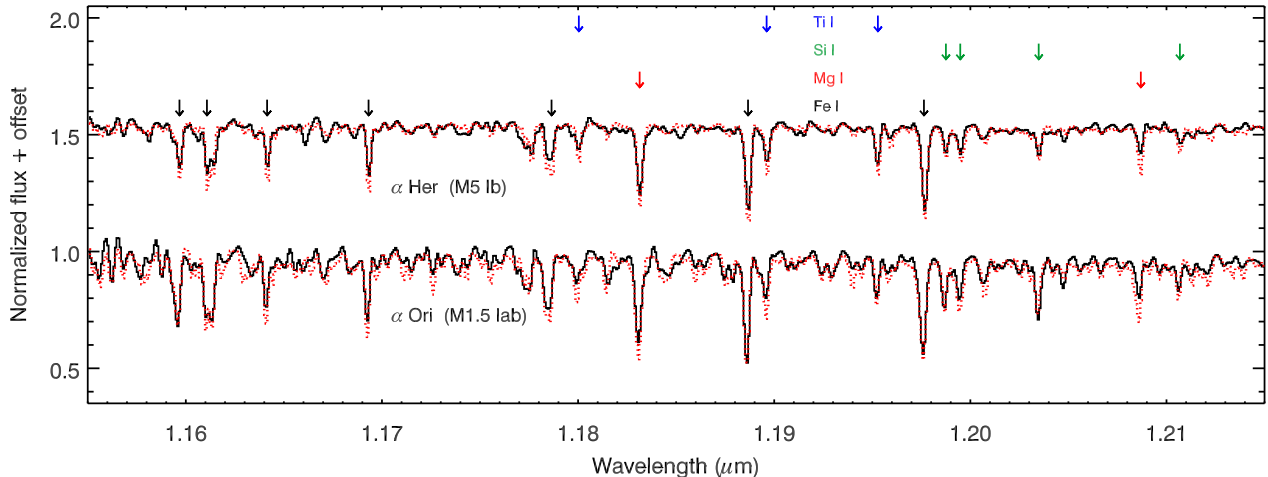}
  \caption{Best fits to the $J$-band spectrum of $\alpha$~Her and
    $\alpha$~Ori, with spectral resolution $R=6000$, published in
    \citet{Wallace00}.  See text for details. }
  \label{fig:r3000}
\end{figure*}

However, at greater spectral resolution the blending of the major
lines is greatly reduced, increasing the number of strong diagnostic
lines that may be studied, as well as increasing the number of
elements that can be probed. This is illustrated in \fig{fig:r3000},
where we plot the spectra of $\alpha$~Her and $\alpha$~Ori originally
presented in \citet{Wallace00}. At this spectral resolution of
$R=6000$ we see the lines of \sii\ and \tii\ as unblended from the
\fei\ lines, particularly at 1.189\um\ and 1.198\um. The results of
the fitting process for these two stars are shown in Table
\ref{tab:r3000}. Note that the abundance measured for $\alpha$~Ori of
\logz=+0.24$\pm$0.22 is entirely consistent with the results of
\citet[][ Fe/H$\approx$+0.1]{Lambert84}

The benefits of this are twofold. Firstly, it is possible to add the
\sii\ and \tii\ lines to the list of `strong features' that are
measured during the fitting process, enabling a more robust fit and
determination of stellar temperature. For example, using only the
\fei\ and \mgi\ lines in the fitting procedure, we measure a
temperature for $\alpha$~Ori of \Teff=3300$\pm$200K. However, these
fits underestimate the strengths of the \sii\ lines, as in the case of
HD~212466 (see Sect.\ \ref{sec:res}). When the \sii\ and \tii\ lines
are folded into the analysis we measure \Teff=3660$\pm$170K, which is
much closer to the temperature appropriate for its spectral type
\citep[M1.5~Iab, \Teff=3710K, using the calibration
  of][]{Levesque05}. We note however that the derived metallicity is
relatively insensitive to this effect, changing by only $\sim$0.1dex.

Secondly, at increased spectral resolution the number of metallic
elements that it is possible to investigate is greater. By studying
the combined abundances of Mg, Si and Ti, combined with more detailed
modelling, it should be possible to make a quantitative comparison of
the relative abundances of Fe-group and $\alpha$-group elements, a key
diagnostic of a galaxy's star-forming history. Though the spectra we
show in \fig{fig:r3000} have resolution of $R=6000$, we have found
that it is still possible to separate the spectral lines at resolution
of only $R=3000$. As lower resolutions mean shorter exposure times for
objects of a given brightness, the limiting distance at which RSGs
could be studied in this way is increased.

\subsection{Extragalactic potential}
\label{sec: potential}

The potential of RSGs as extragalactic abundance probes is enormous. With
absolute magnitudes of $M_{J}\sim -8$ to $-11$ mag and very characteristic
colors ( $V - I \approx 2.0$ mag) they can easily be detected as individual
objects in galaxies at large distances. Using new IR multi-object
spectrographs such as KMOS at the ESO VLT or MOSFIRE at Keck one will be
able to reach limiting magnitudes of $M_{J} \approx 19$ mag with a spectral
resolution of $R$=3000 and a signal-to-noise of 100 in long exposure
sequences (for instance, totalling two nights). In this way quantitative
spectroscopy of RSGs out to 10 Mpc distances will be possible allowing
precise chemical abundances of galaxies to be determined and the evolution
of galaxies to be disentangled.

The future, with next generation of 30m+ ground-based telescopes and
the James Webb Space Telescope, is even more promising. These new
facilities will be optimized for very high spatial resolution
observations in the IR, so that the gain in limiting fluxes for
multi-object spectrographs increases with the fourth power of aperture
diameter. For instance, using the E-ELT exposure time estimator and
including the dramatic effects of adaptive optics we will be able to
go down $M_{J} \approx 22$ mag, which will give us a limiting distance
of 30 Mpc. This opens up a truly remarkable volume of the local
universe for very detailed and precise studies of the formation and
chemical evolution of entire galaxy clusters, such as Virgo, Fornax,
Puppis and Eridanus.


\section{Conclusions} \label{sec:conc}
To demonstrate that it is possible to get accurate and precise
abundance measurements from Red Supergiants (RSGs) at modest spectral
resolutions of $R$=2-3000, we have studied a sample of RSGs in the
Solar neighbourhood in conjunction with the latest MARCS model
atmospheres. With no a priori assumptions about stellar parameters, we
derive chemical abundances for each object which are close to the
Solar value, with a mean abundance for the whole sample of
0.13$\pm$0.14dex. This dispersion in the mean is notwithstanding any
intrinsic spread inherent in the sample, and is indicative of our
experimental uncertainty. The level of precision achieved by this
method is therefore as good as any other abundance indicator. We find
that the derived abundance levels are relatively insensitive to the
other model parameters of effective temperature, microturbulent
velocity, CNO mixture and surface gravity, though the effects of very
low surface gravities (\logg$<$0.0) have not been investigated here.

Though we do find effective temperatures which are on average lower
than the literature values by $\sim$100-200K, this has very little
impact on the derived abundances since \logz\ is extremely insensitive
to \Teff. Forcing the fitted temperature to be the same as the
literature temperature for each star resulted in output abundances
that were very similar and within the errors of the best
fits. Furthermore, we argue that this \Teff\ discrepancy may be solved
at resolutions of $R>3000$ when the blending of key diagnostic lines
is diminished.

The results of our work therefore suggest that it is possible to
obtain accurate abundance information at spectral resolutions which
are much lower than commonly used at present. This opens the door for
abundance studies of entire galaxies at infrared wavelengths, out to
distances of several Mpc.

As for the future, we will shortly present an analysis of a sample of
RSGs from the two Magellanic Clouds, showing that this technique can
be applied across a large range of metallicities. We are also in the
process of expanding the model grid to include a broader range of
gravities, better sampling in $\xi$-space, and with accurate synthetic
spectra computed at high resolution.

\section*{Acknowledgments}
We thank the referee Georges Meynet for useful comments and
suggestions that improved the paper, and Bertrand Plez, Livia Origlia
and Joe Mottram for useful discussions during the course of this
work. We have made extensive use of the MARCS model atmosphere grid,
available at {\tt http://marcs.astro.uu.se}; the IRTF spectral
library, available at {\tt
  http://irtfweb.ifa.hawaii.edu/$\sim$spex/IRTF\_Spectral\_Library};
and the {\sc pollux} database ({\tt
  http://pollux.graal.univ-montp2.fr} ), operated at GRAAL (Université
Montpellier II - CNRS, France) with the support of the PNPS and INSU.

\bibliographystyle{/fat/Data/bibtex/apj}
\bibliography{/fat/Data/bibtex/biblio}

\end{document}